\documentclass[12pt]{amsart} 
\input epsf  
\usepackage[dvips]{graphicx} 
\usepackage{amsmath}
\usepackage{amssymb}
\usepackage{amscd}
\usepackage{bbm}
\usepackage{hyperref}

\begin{document}

\newenvironment{proofof}[1]{\medskip\noindent
   \textbf{Proof of #1:} }{\hfill $\blacksquare$\par\medskip}

\newcommand{\be}{\begin{equation}}
\newcommand{\ee}{\end{equation}}
\newcommand{\ds}{\displaystyle}
\newcommand{\id}[1]{\mathbbm{1}_{#1}}
\newcommand{\iden}{\mathbbm{1}}
\newcommand{\dens}[1]{\langle \eta_{#1} \rangle}
\newcommand{\sigid}{(\sigma \otimes \iden)}
\newcommand{\exi}[1]{\langle \xi_{#1} \rangle}
\newcommand{\aver}[1]{\langle #1 \rangle}
\newcommand{\B}{\mathbb{B}}

\newtheorem{thm}{Theorem}
\newtheorem{cor}[thm]{Corollary}
\newtheorem{lem}[thm]{Lemma}
\newtheorem{defn}{Definition}
\newtheorem{rem}{Remark}
\newtheorem{conj}{Conjecture}

\makeatletter
\def\Ddots{\mathinner{\mkern1mu\raise\p@
\vbox{\kern7\p@\hbox{.}}\mkern2mu
\raise4\p@\hbox{.}\mkern2mu\raise7\p@\hbox{.}\mkern1mu}}
\makeatother

\newcommand{\marginnote}[1]
           {\mbox{}\marginpar{\raggedright\hspace{0pt}{$\blacktriangleright$
#1}}}

\title[Asymmetric Glauber Dynamics]
{Algebraic properties of a Disordered Asymmetric Glauber Model}
\author{Arvind Ayyer}
\email{ayyer@math.ucdavis.edu}
\address{Institut  de Physique Th\'eorique, C. E. A.  Saclay,
 91191 Gif-sur-Yvette Cedex, France}
\address{Department of Mathematics, Mathematical Sciences Building,
  One Shields Ave, University of California, Davis CA 95616}
\date{\today}

\begin{abstract}
We consider an asymmetric variant of disordered Gla\-uber dynamics of
Ising spins on a one-dimensional lattice, where each spin flips
according to the relative state of the spin to its left.  Moreover,
each bond allows for two rates; flips which equalize nearest neighbor
spins, and flips which ``unequalize'' them.  In addition, the leftmost
spin flips depending on the spin at that site.  We explicitly
calculate all eigenvalues of the transition matrix for all system
sizes and conjecture a formula for the normalization factor of the
model. We then analyze two limits of this model, which are analogous
to ferromagnetic and antiferromagnetic behavior in the Ising model for
which we are able to prove an analogous formula for the normalization
factor.
\end{abstract}

\maketitle

\section{Introduction} 
\label{sec:intro}
The Ising model with Glauber dynamics \cite{Glauber} has been an
extremely important source for understanding time dependent behavior
in the Ising model. Moreover, the simplicity of the model has inspired
the creation of a number of related models which have been amenable to
rigorous combinatorial and probabilistic techniques.  Most of the
studies, by far, have been on the probabilistic side, with systems
being studied on the continuum, on infinite lattices in arbitrary
dimension, or considering asymptotics of large finite lattices. The
literature on Glauber dynamics in the Ising model is exceedingly
large. When bonds between neighboring sites have random strengths,
there seem to be a few studies for large or infinite systems,
eg. \cite{falk2,newstein,aliev,crisom}. We know of very few
combinatorial results, eg \cite{falk1, diafanron1} and they correspond
to situations where the bond strengths are not arbitrary. In what
follows, we show that a model with Glauber-like dynamics exhibits rich
algebraic and combinatorial structure.

We consider a simplified one-dimensional version of the Ising model
with spin-flip dynamics, on sites labelled 1 to $L$,
and where there are two kinds of spins, which we label as occupation
numbers $\{0,1\}$. The transition rules are given by an
asymmetric version of the usual rule. Each site $i$ looks only to
the site to its left, $i-1$, and the $i$th site switches its state
with a rate that depends only on whether both sites are in the the
same state or not. If the state of both sites is different, the
transition occurs with rate $\alpha_i$ (Glauber) and if they are the
same, it occurs with rate $\beta_i$ (anti-Glauber).  In other words,
$\alpha_i$'s try to homogenize the system and $\beta_i$'s try to make
nearest neighbors opposite.  In addition there are boundary
interactions at the first site, which flips with rate $\alpha_1$ if it
contains a particle and $\beta_1$ if it does not.

This model can be thought of as a voter model \cite{liggett} within a
hierarchical demographic, that is to say individual $i-1$ influences
the opinion of individual $i$, but not vice versa. The rate $\alpha_i$
governs the tendency of $i-1$ being a usual campaigner, whereas
$\beta_i$ governs the tendency of $i-1$ being a double agent.


This model can also be thought of as a variant of the East model
studied in \cite{alddia1,diafanron1} on the one dimensional
lattice. In the east model, a spin was allowed to flip only if the
site to its left was occupied and the rate of the flip depended on
whether the site was occupied or not.

Crisanti and Sompolinsky \cite{crisom} considered a mean-field model
of Glau\-ber dynamics on Ising spins with asymmetric bonds with random
stren\-gths to understand properties of neural networks. This work is not
related to it, but the idea is similar. In their work, asymmetry
refers to partial asymmetry, instead of the total asymmetry assumed here.

In an earlier work, the steady state of the border process for this
model with $\beta_i=0$ and $\alpha_i=1$ for $i>0$, the asymmetric
annihilation process, was obtained \cite{aymal} by using a transfer
matrix ansatz. The spectrum of the transition matrix for $\alpha_i$
unequal was calculated in \cite{aystr}.  This model, even with
$\beta_i=0$ is a clear generalization because one can distinguish here
between particles and holes, whereas the former could not. In
addition, there is no analog of the rate for $\beta_i$ in the earlier
works because that would have amounted to creation of particles in the
bulk.

We will write down the precise statement of the Markov chain in
Section~\ref{sec:mod}. The main result of the paper is an explicit
formula for the eigenvalues of the Markov chain for any size $L$,
which we prove in Section~\ref{sec:mm}. This leads, in an obvious way,
to a formula for the spectral gap of the system.

We will show that the transition matrix of the model exhibits a simple
recurrence relation allowing one to express the matrix for a system of
size $L$ in terms of that of size $L-1$ in
Section~\ref{sec:recurmm}. We prove a formula for the density in this
disordered model in Section~\ref{sec:dens}. We conjecture a formula
for the normalization factor for the distribution, informally called
the ``partition function'' in Section~\ref{sec:partfn}.  Lastly, we
consider two special cases, which we naturally label the ferromagnetic
and the antiferromagnetic limits in Section~\ref{sec:spec}.

\section{The Model}
 \label{sec:mod}

We consider a  nonequilibrium system on a finite lattice with 
 $L$ sites labelled from 1 to $L$. Each  sites is occupied by one of
 two spins $1$ and $0$.
The evolution rule in the  bulk, for site $i$ depending on the spin at
site $i-1$,  is   given by 
\be \label{bulk}
\begin{split}
&10 \to 11 \; \text{ and } 01 \to 00 \; \text{with rate } \alpha_i \, , \\
&11 \to 10 \; \text{ and } 00 \to 01 \; \text{with rate } \beta_i \, .
\end{split}
\ee

The evolution of the first  site is  given by
\be \label{lbound}
\begin{split}
&1 \to 0 \; \text{with rate } \alpha_1,\\
&0 \to 1 \; \text{with rate } \beta_1. 
\end{split}
\ee

The rule at the left boundary \eqref{lbound} is constructed by
supposing that there is a virtual site labelled 0 to the left of the
first site which is always empty.  The left boundary
conditions are deduced from the bulk rules \eqref{bulk} by looking at
the second component of the bond.

\begin{rem} \label{rem:symm}
  There are two symmetries of the model; the first is manifested by
  interchanging all $\alpha_i$'s and $\beta_i$'s as well as flipping
  spins at all odd sites, and the second, by leaving $\alpha_1$ and
  $\beta_1$ as they are, interchanging all other $\alpha_i$'s and
  $\beta_i$'s and flipping spins at all even sites.
\end{rem}

\begin{rem} \label{rem:corrfn} 
   This model has the property that correlation functions $ \langle
   \eta_{i_1} \dots \eta_{i_n} \rangle$ does not depend on the state
   of $\eta_{i_n+1}$ simply because site $i_n+1$ cannot influence any
   site less than or equal to $i_n$ Further, as is well-known for
   Glauber dynamics, correlation functions of $n$ sites like the one
   above  depend only on those of sites less than or equal to $n$.
\end{rem}

\section{Spectrum of the transition matrices} \label{sec:mm}
We observe that the characteristic polynomial of the transition
matrices factorize into linear factors and has an explicit
formula. This is reminiscent of the formula for the asymmetric
annihilation process conjectured in \cite{aymal} and proved in
\cite{aystr}. Let $\B_L$ be the set of binary vectors of size $L$, that
is vectors of length $L$ whose elements are either 0 or 1.

\begin{thm} \label{thm:charpoly}
The characteristic polynomial of $M_L$ is given by 
\be \label{charpoly}
|M_L - \lambda \id{L}| = 
\prod_{b \in \B_L} \left(\lambda+ \sum_{i=1}^L b_i(\alpha_i+\beta_i)
  \; \right) 
\ee
\end{thm}

For example, when $L=2$, we have
\be
|M_2 - \lambda \id{2}| = 
\lambda\, 
\left( \lambda+\alpha_{{1}}+\beta_{{1}} \right)  
\left( \lambda+\alpha_{{2}}+\beta_{{2}} \right)  
\left( \lambda+\alpha_{{1}}+\alpha_{{2}}+\beta_{{1}}+\beta_{{2}} \right)  
\ee

The proof of Theorem~\ref{thm:charpoly} will turn out to be a simpler
version of the proof of the eigenvalues in the asymmetric annihilation
process \cite{aystr}. We will first define a slight rearrangement of a
Hadamard matrix. Let $\B_L$ be ordered lexicographically. For $b,c \in
\B_L$, define the square matrix $H_L$ of size $2^L$ by
\be
H_L = \frac 1{2^{L/2}}\left( (-1)^{\hat b \cdot c} \; \right)_{b,c \in \B_L},
\ee
where $\hat b$ is the reverse of $b$. 
Note that $H_L$ is symmetric because 
\be
(H_L)_{b,c} = (-1)^{\hat b \cdot c}  = (-1)^{b \cdot \hat c} =
(H_L)_{c,b}.
\ee

This is different from the usual definition in which elements of $H_L$
are written as $(-1)^{b \cdot c}$. For example, the  matrix
for $L=3$ is given by
\be
H_3 = \frac 1{2 \sqrt{2}} 
\left( \begin {array}{cccccccc} 1&1&1&1&1&1&1&1\\
\noalign{\medskip}1&1&1&1&-1&-1&-1&-1\\
\noalign{\medskip}1&1&-1&-1&1&1&-1&-1 \\ 
\noalign{\medskip}1&1&-1&-1&-1&-1&1&1\\
\noalign{\medskip}1&-1&1&-1&1&-1&1&-1\\
\noalign{\medskip}1&-1&1&-1&-1&1&-1&1\\ 
\noalign{\medskip}1&-1&-1&1&1&-1&-1&1\\ 
\noalign{\medskip}1&-1&-1&1&-1&1&1&-1
\end {array} \right).
\ee

\begin{lem} \label{lem:had2}
$H_L^2 = \id{L}$.
\end{lem}

\begin{proof}
We first look at the diagonal terms in $H_L^2$.
\be
(H_L^2)_{b,b} = \sum_{c \in \B_L} (H_L)_{b,c} (H_L)_{c,b} = \frac
1{2^L} \sum_{c \in \B_L} (-1)^{2 \hat b \cdot c} = 1.
\ee

Now suppose $b \neq d$. Then
\be
(H_L^2)_{b,d} = \sum_{c \in \B_L} (H_L)_{b,c} (H_L)_{c,d} = \frac
1{2^L} \sum_{c \in \B_L} (-1)^{ (b+d) \cdot \hat c},
\ee
where the addition of binary vectors is the usual xor-addition:
$x+(1-x)=1$ by definition and $x+x=0$ if $x$ is a bit.

Now, for any fixed $b,d$, we will construct an sign-reversing
involution on $\B_L$. Since $b \neq d$, $b+d$ contains at least one
entry equal to 1. Consider the leftmost such entry, at position $i$,
say. To any element $c$, associate another element
$$c \mapsto c'=(c_1,\dots,c_{L-i},1-c_{L-i+1},c_{L-i+2},\dots,c_L).$$
Clearly, this is an involution and moreover, 
$$(-1)^{ (b+d) \cdot \hat c}+(-1)^{ (b+d) \cdot \hat c'} = 0.$$

We have therefore partitioned $\B_L$, for any fixed $b,d, b
\neq d$, such that  the number of terms which contribute +1 to the sum
is exactly the same as that which contribute -1. Hence $(H_L^2)_{b,d}
= 0$.
\end{proof}

We now write down the transition matrix $M_L$ in this language. We use
the convention that $(M_L)_{b,c}$ represents the transition rate from
$c \to b$, with the end result that $M_L |v_L\rangle = 0$ for the steady state
{\em column} vector $|v_L \rangle$.
\be \label{binmarkov}
(M_L)_{b,c} = \begin{cases}
\alpha_i, & b+c=0 \dots 0\underbrace{1}_i 0\dots 0 \\
 &\text{ and } c_{i-1}+c_i = 1,\\
\beta_i, & b+c=0 \dots 0\underbrace{1}_i 0\dots 0 \\
& \text{ and } c_{i-1}=c_i, \\
\ds -\sum_{\substack{i=1 \\ c_{i-1}+c_i=1}}^L \alpha_i  
-\sum_{\substack{i=1 \\ c_{i-1}=c_i}}^L \beta_i, & b=c, \\
0, & \text{otherwise},
\end{cases}
\ee

\begin{lem}
$H_L M_L H_L$ is lower-triangular.
\end{lem}

\begin{proof}
We will show the result by explicitly demonstrating the structure of
non-zero terms. By definition,
\be
(H_L M_L H_L)_{c,d} = \frac 1{2^L} \sum_{e,f \in \B_L} (-1)^{c \cdot
  \hat e + f \cdot \hat d} (M_L)_{e,f}.
\ee

We will divide the proof into three parts. We first consider diagonal
elements, then certain off-diagonal elements which are non-zero, and
finally show that all other off-diagonal elements are zero by
introducing another sign-reversing involution just as in
Lemma~\ref{lem:had2}.

\begin{itemize}
\item $c=d$.

Then $c \cdot \hat e + f \cdot \hat d = c \cdot(\hat e + \hat
f)$. Suppose $e$ differs from $f$ only at position $i$, then
$(M_L)_{e,f}$ contributes either $\alpha_i$ of $\beta_i$ and the sign
is always the same: $+1$ if $\hat c_i = 0$ and $-1$ otherwise. On the
other hand, if $e=f$, this sum is identically 0 and therefore we get a
contribution of $-\alpha_i$ if $f_{i-1}+f_i=1$ and $-\beta_i$ if
$f_{i-1}=f_i$ from \eqref{binmarkov}. 

To summarize, we get $2^{L-1}$ terms contributing $\pm \alpha_i, \pm
\beta_i$, all of the same sign, from summands $e \neq f$ and $2^{L-1}$
terms contributing $-\alpha_i, -\beta_i$ from summands $e =f$. These
will clearly cancel if $\hat c_i = 0$ and will give
$2^{L-1}(-2\alpha_i-2\beta_i)$ if $\hat c_i = 1$. Thus,
\be \label{diagterm}
(H_L M_L H_L)_{c,c} = -\sum_{i=1}^L \hat c_i (\alpha_i+\beta_i).
\ee

\item $c \neq d$

For now, we look at conditions on $c,d$ such that $(M_L)_{e,f} =
\alpha_i$ and $(-1)^{c \cdot \hat e + f \cdot \hat d}$ always has the
same sign. This happens when $e$ differs from $f$ only
at position $i$ and $f_{i-1}+f_i=1 \Rightarrow e_{i-1}=e_i$. Now,
\be
\begin{split}
c \cdot \hat e + f \cdot \hat d &= e \cdot \hat c + f \cdot \hat d, \\
&= e \cdot (\hat c + \hat d) + (e+f) \cdot \hat d, \\
&= f \cdot (\hat c + \hat d) + (e+f) \cdot \hat c.
\end{split}
\ee Since $e+f$ is fixed, only the former term contributes to the
change of sign. As we vary $e$, this term does not change sign if and
only if $(\hat c + \hat d)_{i-1} = (\hat c + \hat d)_i$ and $c_j =
d_j$ for all other values of $j$. Since we are assuming $c \neq d$,
the only way this can happen is if $\hat c+ \hat d = 0\dots0110\dots0$
with 1s in the $i-1$th and $i$th position. If this is the case, the
sign of all these terms is determined just by $\hat d_i$, or
equivalently, $\hat c_i$. This gives us a total contribution of
$2^{L-1} \alpha_i (-1)^{\hat d_i}$.

We also have contributions of $\alpha_i$ from diagonal terms $e=f$ and
$e_{i-1}+e_i=1$ from \eqref{binmarkov}. In that case, $c \cdot \hat e
+ f \cdot \hat d = e.(\hat c + \hat d)$. Since we just argued that
$(\hat c + \hat d)_{i-1} = (\hat c + \hat d)_i$, diagonal terms always
contribute a term of $(-1)^1 (-\alpha_i)$, giving us a grand total of
$2^{L-1} \alpha_i$.

Therefore, diagonal contributions will cancel off-diagonal ones if
$\hat d_i = 1, \hat c_i=0$ and will add to them if $\hat d_i = 0, \hat
c_i=1$. 

By a very similar reasoning, we will get a contribution of
$-\beta_i$ under exactly the same circumstances. The relation $(\hat c
+ \hat d)_{i-1} = (\hat c + \hat d)_i$ remains the same, but the roles
are of $c$ and $d$ are reversed, giving us an overall negative
sign. 

Assuming all other off-diagonal terms are zero, 
$$(H_L M_L H_L)_{c,d} =\alpha_i - \beta_i$$ 
if and only if $c$ and $d$ differ only in the
$i-1$th and $i$th position and moreover $d_i=0$ and $c_i=1$. Therefore
$d < c$ in the lexicographic order and the matrix is lower triangular.

\bigskip
\item Involution.

To complete the proof, we have to show that when $c$ and $d$ do not
satisfy $\hat c+ \hat d = 0\dots0110\dots0$ with 1s in the $i-1$th and
$i$th position, $(H_L M_L H_L)_{c,d} =0$. To do this, we introduce
another sign-reversing involution just as in Lemma~\ref{lem:had2}. For
a fixed $c,d$ and any pair $(e,f)$ which gives a contribution of
$\alpha$ to $(H_L M_L H_L)_{c,d}$, say, we find another pair $(e',f')$
which also give the same contribution but such that $c \cdot \hat e + f \cdot
\hat d = c \cdot \hat e' + f' \cdot \hat d +1$. Notice that $(e,f)$
satisfies the condition that $\hat e$ and $\hat f$ differ only in the
$i$th position and that $f_{i-1}+f_i = 1$.

This is easily done. Since $c \neq d$, find the smallest value of $j$,
$j \neq i, i-1$ such that $\hat c_j + \hat d_j =1$. Then choose
$(e',f')$ as
\be
\begin{split}
e \mapsto& e'=(e_1,\dots,e_{L-j},1-e_{L-j+1},e_{L-j+2},\dots,e_L), \\
f \mapsto& f'=(f_1,\dots,f_{L-j},1-f_{L-j+1},f_{L-j+2},\dots,f_L).
\end{split}
\ee
Then $e+f=e'+f'$ and $e.(\hat c + \hat d) = e'.(\hat c + \hat d) \pm
1$ and we are done. Notice that we need $j \neq i-1,i$ to ensure that
$e_{i-1}=e_i$ remains intact. If $\hat c$ differs from $\hat d$ only
at these two positions, then we already proved that there can be no
such involution.
\end{itemize}

\end{proof}

We have now done all the work necessary to write down a proof of the
main result.

\begin{proofof}{Theorem~\ref{thm:charpoly}}
Since $H_L$ is symmetric and $H_L^2=1$ by Lemma~\ref{lem:had2}, 
$H_L M_L H_L$ has the same characteristic polynomial as $M_L$. But the
former is lower triangular. The characteristic polynomial is therefore
\be
| M_L - \lambda \id{L}| = \prod_{i=1}^{2^L} (-\lambda - (H_L M_L
H_L)_{i,i}).
\ee
Using the representation of the rows and columns using boolean vectors
of size $L$ and using \eqref{diagterm}, we have the required result.
\end{proofof}


\begin{cor}
The spectral gap for the system of size $L$ is given by
\be
\min_{i=1}^L(\alpha_i+\beta_i).
\ee
\end{cor}

\begin{proof}
The proof directly follows from the explicit construction of
eigenvalues in Theorem~\ref{thm:charpoly}.
\end{proof}

\section{Recurrence Relation for the Transition
  Matrices} \label{sec:recurmm}
We will prove that there is a recursion of order one
among the transition matrices. Let $M_L(\alpha_1, \dots, \alpha_L;
\beta_1, \dots, \beta_L)$ represent the transition matrix for the
system of size $L$ with parameters $\alpha_i$ and $\beta_i$.

\begin{thm} \label{thm:mm}
Let $\iden$ denote the identity matrix of size $2^{L-1}$. Then 
$M_L$ can be expressed in $2 \times 2$ block-diagonal form as
\be \label{mmdecomp}
M_{L}(\alpha_1, \dots, \alpha_L;
\beta_1, \dots, \beta_L) 
= \left(\begin{array}{c|c}
    M_{1,1}& \alpha_1 \iden \\
\hline
  \beta_1 \iden & M_{2,2}
  \end{array} \right),
\ee
where
\be
\begin{split}
M_{1,1} &= M_{L-1}(\alpha_2, \alpha_3,\dots, \alpha_{L};\beta_2,\beta_3,
  \dots, \beta_{L}) -\beta_1 \iden, \\
M_{2,2} &= M_{L-1}(\beta_2, \alpha_3,\dots, \alpha_{L};\alpha_2,\beta_3,
  \dots, \beta_{L}) -\alpha_1 \iden,
\end{split}
\ee
with the initial matrix for $L=1$ given
by
\be \label{mmic}
M_1(\alpha_1;\beta_1) = \begin{pmatrix}
  -\beta_1 & \alpha_1 \\
  \beta_1 & -\alpha_1
  \end{pmatrix}.
\ee
\end{thm}

\begin{proof}
The matrix for size 1 given in \eqref{mmic} is easy to check.
The matrix recursion \eqref{mmdecomp} is proved by looking at transitions that
binary vectors beginning with different two-bit vectors undergo.  Let
$v,v'$ denote binary vectors of length $L$ and $w,w'$ denote vectors
of length $L-1$. There are then four possibilities of transitions
depending on the first bits of $v$ and $v'$, namely $v=0w$ or $v=1w$
and similarly for $v'$.
\begin{enumerate}
\item $0w \to 1w'$ (resp.  $1w \to 0w'$): \\
  These transitions are only possible if $w=w'$ and then occurs with rate
  $\beta_1$ (resp. $\alpha_1$). This is why the $(2,1)-$th
  (resp. $(1,2)-$th) block of $M_L$ is $\beta_1 \iden$ (resp. $\alpha_1 \iden$).

\item $0w \to 0w'$ (resp. $1w \to 1w'$): \\   
  In this case, all the transitions for the system of size $L-1$ go
  through but with rates whose indices are one more than that of the
  smaller system simply because one extra site has been added to the
  left. For example, what occurred with rate $\alpha_2$ in the system
  of size $L-1$ occurs with rate $\alpha_3$ now. Because of the extra
  transitions in the off-diagonal blocks (described above), we must
  add negative terms to the diagonal so that column sums are
  zero. Therefore we subtract $\beta_1 \iden$ (resp. $\alpha_1 \iden$)
  from $M_{1,1}$ (resp. $M_{2,2}$).
\end{enumerate}

\end{proof}

For later consideration, we also express the recursion relation using
a second order recurrence relation. Expressing the transition matrix
for the system of size $L-1$ as
\be \label{mmdecomp2}
\begin{split}
&M_{L}(\alpha_1, \dots, \alpha_L;
\beta_1, \dots, \beta_L) 
= \\
&\left(\begin{array}{c|c|c|c}
    M_{1,1}- & \alpha_2 \iden & \alpha_1 \iden & 0\\
(\beta_1+\beta_2)\iden & & &\\
\hline
  \beta_2 \iden & M_{2,2}-  & 0 & \alpha_1
  \iden \\ 
& (\beta_1+\alpha_2)\iden & & \\
\hline
  \beta_1 \iden & 0 & M_{1,1}-   & \beta_2
  \iden \\ 
& & (\alpha_1+\alpha_2)\iden & \\
\hline
  0 & \beta_1 \iden &  \alpha_2  \iden & M_{2,2}-  \\
& & &  (\alpha_1+\beta_2)\iden 
  \end{array} \right),
\end{split}
\ee
where
\be
\begin{split}
M_{1,1} &= M_{L-2}(\alpha_3,\alpha_4,\dots, \alpha_{L};\beta_3,\beta_4,
  \dots, \beta_{L}),  \\
M_{2,2} &= M_{L-2}(\beta_3, \alpha_4,\dots, \alpha_{L};\alpha_3,\beta_4,
  \dots, \beta_{L}).
\end{split}
\ee

\begin{rem}
Notice that the transition matrix $M_L$ is invariant under the
transformation $M_L \leftrightarrow (M_L)^T$ and $\alpha
\leftrightarrow \beta$. 
\end{rem}

\section{Density in the Steady State} \label{sec:dens}
We compute the density of $1$'s in the steady state at site $k$ in a
system of size larger than $k$. Using Remark~\ref{rem:corrfn}, this
density is completely independent of the size of the system.

Let $S_k$ be the set of subsets of $[k] = \{1,\dots,k\}$ with an odd
number of elements, and for $s \in S_k$, let $\bar s = S_k \setminus s$.

\begin{lem}
The density of $1$'s at site $k$ is given by
\be \label{densk}
\langle \eta_k \rangle = \frac{\ds \sum_{s \in S_k} \prod_{i \in s}
  \beta_i \prod_{j \in \bar s} \alpha_j}{\ds \prod_{j=1}^k
  (\alpha_j+\beta_j)}.
\ee
\end{lem}

\begin{proof}
The master equation for the density of $1$'s at site 1 is
\be
\frac d{dt} \dens{1} = \beta_1 \aver{1-\eta_1} - \alpha_1 \dens{1}=0,
\ee
which implies $\eta_1 = \frac{\beta_1}{(\alpha_1+\beta_1)}$, which is
consistent with the \eqref{densk} with $k=1$.

For $k>1$, the master equation is 
\be
\begin{split}
\frac d{dt} \dens{k} = \beta_k &\aver{(1-\eta_{k-1})(1-\eta_k)} 
+ \alpha_k \aver{\eta_{k-1}(1-\eta_k)} \\
&- \alpha_k\aver{(1-\eta_{k-1})\eta_k} - \beta_k \aver{\eta_{k-1}
  \eta_k}=0,
\end{split}
\ee
which on simplifying gives
\be
\dens{k} = \frac{\alpha_k \dens{k-1} + \beta_k \aver{1-\eta_{k-1}}}
{\alpha_k + \beta_k}.
\ee
One can check that \eqref{densk} satisfies this recurrence.
\end{proof}

\section{The Normalization Factor} \label{sec:partfn}
Consider the system of size $L$. For any configuration $\eta$, we can
express its steady state probability as a numerator divided by a
denominator.  We define the normalization factor $Z_L(\vec \alpha,\vec
\beta)$ as the least common multiple of the denominators of the steady
state probabilities for all configurations $\eta$ such that the
greatest common divisor of the corresponding numerators is 1.

We have been able to find a conjecture for $Z_L$ but have not been
able to prove the formula in general. In certain special cases
discussed in Section~\ref{sec:spec}, the following conjecture does
reduce correctly.

\begin{conj}
The normalization factor for a system of size $L$ with rates
$\alpha_i$ and $\beta_i$ is given by
\be
Z_L(\vec \alpha, \vec \beta) = \prod_{i=1}^L (\alpha_i + \beta_i)
\prod_{1 \leq i < j \leq L} (\alpha_i+\beta_i+\alpha_j+\beta_j).
\ee
\end{conj}

\section{Special Cases: Ferromagnetic and Antiferromagnetic models} 
\label{sec:spec}

We now analyze two limits of this model which depend on two
parameters, $\alpha_1 = \alpha$ and $\beta_1 = \beta$. Both these
limits can be interpreted as nonequilibrium analogs of the
ferromagnetic and antiferromagnetic Ising models respectively, hence
the terminology. 

The strategy for the analysis of both cases is similar. We will first
use Theorem~\ref{thm:mm} to write a first order recurrence for the
transition matrices.  We will then show that the transfer matrix
ansatz  \cite{aymal} holds, using which we will express the steady
state probabilities of the system of size $L$ in terms of the system
of size $L-1$. 

We first recall the definition of the transfer matrix ansatz. 
A family of  Markov processes  satisfies the 
transfer matrix ansatz if there exist matrices $T_{L}$ for all sizes $L$
such that 
\be \label{tm}
M_{L+1} T_{L} = T_{L} M_L \, .
\ee
 We also  impose that  this equality is nontrivial in the sense that
\be
M_{L+1} T_{L} \neq 0. \label{cond:nontriv}
\ee
The rectangular transfer matrices or conjugation matrices $T_{L}$ can
be interpreted as a consequence of the integrability of the model. 

For the system we consider here, the transition matrices are of size
$2^L$. It is most convenient for us to take the naturally ordered
basis of binary sequences of size $L$.  For example, when $L=2$, the
ordered list is $(00,01,10,11)$.

Using the transfer matrices, we will  calculate the
normalization factor $Z_L$ for both these models.
Letting $|v_1 \rangle$ be the Perron-Frobenius eigenvector
of the system of size one, we can define
\be \label{ztm}
Z_L = \langle 1_L| T_{L-1} T_{L-2} \cdots T_1 |v_1 \rangle,
\ee
where $\langle 1_L|$ is the row vector of size $2^L$ composed of 1's.

For these special cases, we will explicitly calculate the density, the
recurrences for the transition matrix and the transfer matrix, as well
as the formula for the normalization factor. We will omit all the
proofs since they are quite straightforward, if somewhat tedious.
The recurrences for the transition matrix follow from the recurrence
\eqref{mmdecomp2} and the others can be provided without too much
difficulty using induction. The technique of proof is the same as was
used in \cite{aymal}.

\subsection{Ferromagnetic limit} \label{sec:fer}
We consider the case $\alpha_i = 1$ and $\beta_i = 0$ for
$i>1$. The densities are easily calculated. The
only term that contributes to the sum in \eqref{densk} is the
subset $\{1\} \in S_k$. Therefore,
\be
\dens{k} = \frac{\beta}{\alpha+\beta},
\ee
for all $k$. Therefore all densities are identical, which is
reminiscent of  the ferromagnetic limit where all spins prefer to be
aligned in the same direction.

\begin{cor} \label{thm:mmfer}
Let $\iden$ denote the identity matrix of size $2^{L-2}$. Suppose
the transition matrix for size $L-1$ is written in block-diagonal form as
\be \label{mmdecompfer}
M^{(F)}_{L-1} = \left(\begin{array}{c|c}
  M^{(F)}_{11} & \alpha \iden \\
\hline
  \beta \iden & M^{(F)}_{22}
  \end{array} \right).
\ee
Then 
\be \label{conjmmfer}
M^{(F)}_L = \left( \begin{array}{c|c|c|c}
M^{(F)}_{11} & \iden & \alpha \iden& 0 \\
\hline
0 & M^{(F)}_{22}-  & 0 & \alpha \iden \\
 &  (1-\alpha+\beta)\iden & & \\
\hline
\beta \iden & 0 & M^{(F)}_{11} -  & 0 \\
 & & (1-\beta+\alpha) \iden & \\
\hline
0 & \beta \iden & \iden & M^{(F)}_{22}
\end{array} \right),
\ee
where $M^{(F)}_L$ is written as a $2 \times 2$ block matrix with each block
made up of matrices of size $2^{L-1}$. The initial matrix for
 $L=1$ is given
by
\be \label{mmicferr}
M^{(F)}_1 = \begin{pmatrix}
  -\beta & \alpha \\
  \beta & -\alpha
  \end{pmatrix}.
\ee
\end{cor}

Let $\sigma_L$ denote the matrix of size $2^L$ with 1's on the
antidiagonal and zeros everywhere else. For example,
\be
 \sigma_1 = \begin{pmatrix}
  0 & 1 \\
  1 & 0
  \end{pmatrix}.
\ee

\begin{lem} \label{lem:tmfer}
Define rectangular matrices $T^{(F)}_L$ of size $2^{L+1} \times 2^L$ using
the following recurrence relation. 
\be \label{tmicfer}
T^{(F)}_{1} = \left( \begin{array}{c c}
1 - \beta + \alpha & \alpha \\
0 & \alpha \\
\beta & 0 \\
\beta & 1-\alpha+\beta
\end{array} \right),
\ee
and define $T^{(F)}_L$ by the following recursion. If
\be \label{tmdecompfer}
T^{(F)}_{L-1} = \left(\begin{array}{c|c}
  T^{(F)}_{11} & T^{(F)}_{12} \\
\hline
  0 & T^{(F)}_{22} \\
\hline
  T^{(F)}_{31} & 0 \\
\hline
  T^{(F)}_{41} & T^{(F)}_{42} 
  \end{array} \right).
\ee
then 
\be \label{conjtmfer}
T^{(F)}_{L} =   \left( \begin{array}{c|c|c|c}
2T^{(F)}_{11} & T^{(F)}_{11} & 2T^{(F)}_{12} & T^{(F)}_{12} \\
\hline
0 & \sigma_{L-2}T^{(F)}_{11} \sigma_{L-2} & 0 & T^{(F)}_{22} \\
\hline
0 & 0 & T^{(F)}_{12} & 0 \\
\hline
0 & 0 & T^{(F)}_{22} & 2 T^{(F)}_{22} \\
\hline
2T^{(F)}_{31} & T^{(F)}_{31} & 0 & 0 \\
\hline
0 & T^{(F)}_{41} & 0 & 0 \\
\hline
T^{(F)}_{31} & 0 & \sigma_{L-2}T^{(F)}_{42} \sigma_{L-2} & 0 \\
\hline
T^{(F)}_{41} & 2T^{(F)}_{41} & T^{(F)}_{42} & 2T^{(F)}_{42}
\end{array} \right). 
\ee 
Then the family of Markov processes given by the transition matrices
$\{M^{(F)}_L\}$ satisfy the transfer matrix ansatz with
$T^{(F)}_L$ defined above, ie $M^{(F)}_{L} T^{(F)}_{L-1} =
T^{(F)}_{L-1} M^{(F)}_{L-1}$. 
\end{lem}

Using Lemma~\ref{lem:tmfer}, we can compute the normalization
factor $Z^{(F)}_L(\alpha,\beta)$ using \eqref{ztm} by setting 
\be \label{v1}
|v_1 \rangle = \begin{pmatrix}
\beta\\
\alpha 
\end{pmatrix}.
\ee
For example, $Z^{(F)}_1 = \alpha + \beta$. We find a remarkable property of the
partition function of the system, namely its super-extensive growth with 
the size of the system.

\begin{cor} \label{cor:denfer}
The partition function of the system of size $L$ is
given by
\be  \label{zeqfer}
Z^{(F)}_L = 2^{\binom{L-1}{2}} (\alpha+\beta) (1+ \alpha+\beta)^{L-1}.
\ee
\end{cor}

\subsection{Antiferromagnetic limit} \label{sec:antiferr}
We consider the case $\alpha_i = 0$ and $\beta_i = 1$ for
$i>1$. Now the
only terms that can contribute to the sum in \eqref{densk} are the
subset $\bar s = \phi,\{1\} \in S_k$. Therefore, the density depends on the
parity of the site. If $k$ is odd (resp. even), $s = [k]$ (resp. $s =
\{2,\dots,k\}$) is the only contribution,
\be
\dens{k} = \begin{cases}
\frac{\beta}{\alpha+\beta}, & k \text{ odd}, \\
\frac{\alpha}{\alpha+\beta}, & k \text{ even}. 
\end{cases}
\ee
This is reminiscent of the antiferromagnetic Ising model in which
neighboring sites prefer to be aligned opposite to one another.

\begin{cor} \label{thm:mmantifer}
Let $\iden$ denote the identity matrix of size $2^{L-2}$. Suppose
the transition matrix for size $L-1$ is written in block-diagonal form as
\be \label{mmdecompantifer}
M^{(A)}_{L-1} = \left(\begin{array}{c|c}
  M^{(A)}_{11} & \alpha \iden \\
\hline
  \beta \iden & M^{(A)}_{22}
  \end{array} \right).
\ee
Then 
\be \label{conjmmantiferr}
M^{(A)}_L = \left( \begin{array}{c|c|c|c}
M^{(A)}_{11}-\iden & 0 & \alpha \iden& 0 \\
\hline
\iden & M^{(A)}_{22} -  & 0 & \alpha \iden \\
 & (\alpha-\beta)\iden & & \\
\hline
\beta \iden & 0 & M^{(A)}_{11} -  & \iden \\
 & & (\beta-\alpha) \iden & \\
\hline
0 & \beta \iden & 0 & M^{(A)}_{22}-\iden
\end{array} \right),
\ee
where $M^{(A)}_L$ is written as a $2 \times 2$ block matrix with each block
made up of matrices of size $2^{L-1}$. The initial matrix for
 $L=1$ is given
by
\be \label{mmicantiferr}
M^{(A)}_1 = \begin{pmatrix}
  -\beta & \alpha \\
  \beta & -\alpha
  \end{pmatrix}.
\ee
\end{cor}

\begin{lem} \label{lem:tmantifer}
Define rectangular matrices $T^{(A)}_L$ of size $2^{L+1} \times 2^L$ using
the following recurrence relation. 
\be \label{tmicantifer}
T^{(A)}_{1} = \left( \begin{array}{c c}
0 & \alpha \\
1 - \beta + \alpha & \alpha \\
\beta & 1-\alpha+\beta \\
\beta & 0 
\end{array} \right),
\ee
and if
\be \label{tmdecompantifer}
T^{(A)}_{L-1} = \left(\begin{array}{c|c}
  0 & T^{(A)}_{12} \\
\hline
  T^{(A)}_{21} & T^{(A)}_{22} \\
\hline
  T^{(A)}_{31} & T^{(A)}_{32} \\
\hline
  T^{(A)}_{41} & 0 
  \end{array} \right).
\ee
then 
\be \label{conjtmantifer}
T^{(A)}_{L} =   \left( \begin{array}{c|c|c|c}
0 & 0 & 0 & T^{(A)}_{12} \\
\hline
0 & 0 & 2T^{(A)}_{22} & T^{(A)}_{22} \\
\hline
\sigma T^{(A)}_{21} \sigma & 2\sigma T^{(A)}_{21} \sigma & T^{(A)}_{12} & 2T^{(A)}_{12}\\
\hline
T^{(A)}_{21} & 0 & T^{(A)}_{22} & 0 \\
\hline
0 & T^{(A)}_{31} & 0 & T^{(A)}_{32} \\
\hline
2T^{(A)}_{41} & T^{(A)}_{41} & 2\sigma T^{(A)}_{32} \sigma & 
\sigma T^{(A)}_{32} \sigma \\
\hline
T^{(A)}_{31} & 2T^{(A)}_{31} & 0 & 0 \\
\hline
T^{(A)}_{41} & 0 & 0 & 0
\end{array} \right). 
\ee 
Then the family of Markov processes given by the transition matrices
$\{M^{(A)}_L\}$ satisfy the transfer matrix ansatz with
$T^{(A)}_L$ defined above, ie 
$M^{(A)}_{L} T^{(A)}_{L-1}  = T^{(A)}_{L-1} M^{(A)}_{L-1}$.
\end{lem}

Using Lemma~\ref{lem:tmfer}, we can compute the normalization factor
$Z^{(A)}_L(\alpha,\beta)$ using \eqref{ztm} by using the same formula for $
|v_1 \rangle$ as in the ferromagnetic case, \eqref{v1}.  We find that
the normalization factor is exactly the same as for the ferromagnetic
model.

\begin{cor} \label{cor:denantifer}
The partition function of the system of size $L$ is
given by
\be  \label{zeqantifer}
Z^{(A)}_L = 2^{\binom{L-1}{2}} (\alpha+\beta) (1+ \alpha+\beta)^{L-1}.
\ee
\end{cor}

\subsection{Relation between these special cases}
The transition matrix $M^{(X)}_L$ for $X=F,A$ is invariant, in addition to the
transposition symmetry, under the transformation 
\be
(M^{(X)}_L)_{i,j} \leftrightarrow (M^{(X)}_L)_{2^L+1-i,2^L+1-j} \text{ and  }
\alpha \leftrightarrow \beta 
\ee
in the usual matrix coordinate notation where $i,j$ vary from 1
to $2^L$.  Similarly, the transfer matrix $T^{(X)}_L$ is invariant under the
transformation 
\be
(T^{(X)}_L)_{i,j} \leftrightarrow (T^{(X)}_L)_{2^{L+1}+1-i,2^L+1-j} \text{ and  } 
\alpha \leftrightarrow \beta
\ee
 where this time $i$ varies from 1 to $2^{L+1}$ and $j$ varies from 1 to
$2^L$.
The normalization factors for the two systems are exactly the same,
$Z^{(F)}_L=Z^{(A)}_L$. One also sees that the recurrence relations for
the transition matrices and the transfer matrices are similar.

The reason that there are so many similarities between the two models
is that one can establish an exact correspondence between the ferromagnetic
and antiferromagnetic limits defined previously using
Remark~\ref{rem:symm}. In this case, we use the second part of the
remark. As a consequence of this coupling, time-dependent correlation
functions in the ferromagnetic model are related to those in the
antiferromagnetic model.

\section*{Acknowledgements}
We thank Paul Krapivsky, Kirone Mallick and Volker Strehl for
discussions. We also thank members of IPhT, CEA Saclay for their
hospitality during the author's unforeseen extended stay.

\end{document}